\documentclass[sigconf]{acmart}

\usepackage{subcaption}
\usepackage{graphicx}
\usepackage{multirow}
\usepackage{csquotes}

\AtBeginDocument{%
  \providecommand\BibTeX{{%
    \normalfont B\kern-0.5em{\scshape i\kern-0.25em b}\kern-0.8em\TeX}}}

\setcopyright{none}
\renewcommand\footnotetextcopyrightpermission[1]{} %
\newcommand{\todo}[1]{}
\renewcommand{\todo}[1]{{\color{red} TODO: {#1}}}

\begin{document}

\title[Gender bias in Indian Journalist-Politician Interactions]{Uncovering Gender Bias within Journalist-Politician\\ Interaction in Indian Twitter}

\author{Brisha Jain}
\affiliation{%
  \institution{Independent researcher}
  \country{India}}
\email{brishajain02@gmail.com}

\author{Mainack Mondal}
\affiliation{%
  \institution{IIT Kharagpur}
  \country{India}}
\email{mainack@cse.iitkgp.ac.in}

\renewcommand{\shortauthors}{Jain et al.}

\begin{abstract}
Gender bias in political discourse is a significant problem on today's social media. Previous studies found that the gender of politicians indeed influences the content directed towards them by the general public. However, these works are particularly focused on the global north, which represents individualistic culture. Furthermore, they did not address whether there is gender bias even within the interaction between popular journalists and politicians in the global south. These understudied journalist-politician interactions are important (more so in collectivistic cultures like the global south) as they can significantly affect public sentiment and help set gender-biased social norms. In this work, using large-scale data from Indian Twitter we address this research gap.

Specifically, we curated a gender-balanced set of the 100 most-followed Indian journalists on Twitter and the 100 most-followed politicians. Then we collected 21,188 unique tweets posted by these journalists that mentioned these politicians. Our analysis revealed that there is a significant gender bias---the frequency with which journalists mention male politicians vs. how frequently they mention female politicians is statistically significantly different ($p<<0.05$). In fact, median tweets from female journalists mentioning female politicians received ten times fewer likes than median tweets from female journalists mentioning male politicians. However, when we analyzed tweet content, our emotion score analysis and topic modeling analysis did not reveal any significant difference between the journalists' tweets mentioning male politicians and those mentioning female politicians. Finally, we found a potential reason for the significant gender bias: the number of popular male Indian politicians is almost twice as large as the number of popular female Indian politicians, which might have resulted in the male-biased popularity of tweets (and even the frequency of receiving tweets). We conclude by discussing the implications of this work for the need for gender diversity in the political discourse of Indian social media and the future development of recommender systems for social media that can address this need.

\end{abstract}

\maketitle

\section{Introduction}

\begin{displayquote}
    \textit{``Since Smiriti Irani is displaying her acting talent in the emotional /angry category, she may `forget ' she isn't on set \& break into a dance''}\footnote{Smriti Irani is a female Indian politician. The original tweet mentioning her is available at \url{https://twitter.com/tehseenp/status/702491795079364609}}
\end{displayquote}

\noindent The tweet above, posted by a male spokesperson of a national political party against a female cabinet minister of India offers a glimpse into the endemic nature of gender bias in political conversations on Indian social media sites like Twitter~\footnote{At the time of performing this study in 2023, the platform was called Twitter rather than x.com. So, we will call our platform Twitter in this paper}. In fact, according to an Amnesty study, one in seven tweets directed towards female politicians in India are abusive in nature~\cite{gender-bias-india}. 

Recent studies have examined gender bias in political discourse on social media~\cite{paper-number-25-mertens-websci, paper-number-999-agarwal-sastry,paper-number-1-AalbergJenssen+2017+17+32}. These studies have shown that the gender of politicians does not influence their own social media posts, but their gender does influence the content directed towards them. However, these studies have two serious shortcomings. 

First, many of these studies exclusively looked into global north and discounted global south. However, this is particularly concerning on several counts. One, Southeast Asia has roughly 527 million active social media users, larger than any other contiguous geography. Of these, roughly 470 million are in India. Also, according to political science literature, southeast Asia has not yet matched the Western Hemisphere in gender equality. Therefore, gender bias in political discourse on social media has not yet been studied for a large group of users active in a less progressive social setting. In this study, we focus on the interactions with politicians from a large country in the global south---India. In India the fact that female politicians face handicaps in their ability to participate in the democratic process has been widely recognised. In India, legislators have sought to address it by passing the Women’s Reservation bill which allocates 33\% of representation in central and state legislative bodies for women. While this bill will provide representation to women, it remains unclear if it will provide women with the voice to initiate and shape political discourse in India. This is because gender bias in social media interactions in less progressive democracies like India is problematic at several levels. It dissuades female politicians from being able to harness the power of social media for political purposes and therefore puts them at a disadvantage compared to their male peers. This can undermine their effectiveness and career prospects. It also marginalizes their views from the political discourse on social media thereby reducing the diversity of opinions available to neutral observers. In our work, we aim to uncover the potential of gender bias in the specific context of digital conversations towards Indian politicians. 

Second, previous studies often focused on the interactions between politicians and the general public. Although this understanding is quite valuable, we noted that there is not much work on specific interaction between politicians and more influential social media users. These interactions are important as they can significantly affect the public sentiment. In fact, recent work by Shekhawat et al. has shown that use of Twitter by Indian politicians to engage with newspapers and influencers via social media has been growing~\cite{shekhawat2021twitter}. The fact that politicians have been keen to engage with influencers seems to suggest that these influencers play a significant role in shaping political discourse on Twitter in India. It is therefore important to examine if even these journalist-politician interactions are suffering from gender bias. However, no previous study has checked if even the interaction between the well-known journalists and popular politicians hints at existing gender bias. 

To that end, we investigate if a gender bias exists in the journalist-politician interactions (i.e., the Indian journalists' social media posts directed towards Indian politicians) within Indian demographics of a popular social media site. We chose Twitter as our experiential platform since it is the most popular medium for political interactions between politicians and voters in India. India has 26.5 million Twitter users, and it is used extensively by politicians to engage with voters, influencers, and legacy media outlets. Specifically, in this study we ask the following two research questions:

    \begin{enumerate}
        \item [\textbf{RQ1:}] Is there gender bias in interaction frequency and popularity of journalist-politician interactions? 
        \item [\textbf{RQ2:}]: Is there gender bias within the content of journalist-politician tweets? 
    \end{enumerate}

The study investigates these questions by curating a gender-balanced dataset of Twitter accounts (using programmatic data collection from Twitter and gender detection) comprising hundred popular (by number of Twitter followers) Indian politicians and hundred popular Indian journalists. Then this study programmatically collected all the tweets posted by these journalists' accounts mentioning the accounts of popular politicians in our dataset. we divided our collected tweets four categories according to the gender of the senders/receivers---Male journalist’s tweets mentioning Male Politicians (\textbf{MJ-MP}), Female journalist’s tweets mentioning Male Politicians (\textbf{FJ-MP}), Male journalist’s tweets mentioning Female Politicians (\textbf{MJ-FP}) and Female journalist’s mentioning Female Politicians (\textbf{FJ-FP}). In total we collected 21,188 unique tweets across these four categories. 

Our analysis revealed a significant gender bias---there are statistically significant differences ($p<<0.05$) in how frequently Male/Female journalists mention Male politicians vs. how frequently they mention Female politicians. Moreover, we found that the Tweets mentioning Male politicians are more popular. In fact, across all metrics (retweets, likes, replies) the median popularity of the posts mentioning female politicians are consistently lower than posts mentioning male politicians. E.g., the median number of likes received by a post in FJ-FP (i.e., posted by female journalist mentioning female politician) is only 35, whereas the median number of likes received by a post in FJ-MP (i.e., posted by female journalist mentioning male politician) is 398---more than ten times higher. Note that these likes are given by the general public. 

However, interestingly, the content analysis revealed no bias---the content of actual tweets posted by journalists towards these politicians is not gender-biased in terms of either emotion scores or the topic of tweets. To that end, we identified a more fundamental reason for the low popularity of tweets towards female politicians---the inherent gender bias in Indian Twitter (and perhaps reflective of Indian society), where  popular male politicians (by number of Twitter followers) are almost twice as many compared to popular female politicians. We conclude this paper, by discussing the limitations of our study and identifying how our findings hints towards the necessity of countering gender bias. We surmise that our work will help Indian social media platform developers towards making systematic algorithmic changes within their recommender systems for countering the ill effects of gender bias observed in this work. 

In the rest of the paper we first present the related work in Section~\ref{sec:related}, describe our data collection strategy (Section~\ref{sec:data}) and analysis methodology (Section~\ref{sec:meth}). Then we present our results and examine the research questions in Section~\ref{sec:results}. Then we present the limitations of our study(Section~\ref{sec:lim}) and conclude (Section~\ref{sec:conclu}).

\section{Related work}\label{sec:related}

\noindent Gender bias in Indian society is well documented both in social and professional settings. Women face discrimination and harassment in Indian corporations [20]. Stereotypical attitudes towards women lead to bias in the Indian judicial system~\cite{paper-number-19-bhowmick}. This bias is so firmly entrenched in societal attitudes towards women that popular cinema (colloquially referred to as Bollywood) panders to and amplifies these biases~\cite{paper-number-21-khadilkar}. Naturally, these attitudes have also spilled over to the political domain. Female politicians face structural bias due to their gender and their path to power is distinct and more challenging than their male peers~\cite{paper-number-23-spary}. Indian journalists are also not immune to social conditioning based on gender. Research shows that Indian journalists who are made aware of their subconscious bias against women behave differently on issues related to women compared to a control group of journalists who receive no such training~\cite{paper-number-23-spary}. Given the outsized role that politics plays in shaping societal mores and the influence of journalists in creating political narratives in a democracy like India, it becomes critical to examine the impact of gender bias in journalists against female politicians. This paper tries to examine this question in detail using data from interactions between journalists and politicians on Twitter.

Extensive research has examined the interaction dynamics between politicians and journalists in various contexts~\cite{paper-number-999-agarwal-sastry,paper-number-1-AalbergJenssen+2017+17+32,paper-number-3-bauer,paper-number-5-Ditonto,paper-number-7-dolan,paper-number-9-evans,paper-number-11-huddy}. However, there has been a notable lack of analysis through the lens of Indian demographics. Twitter has become a pivotal platform for political discourse and news dissemination in India, making it essential to understand how politicians and journalists interact on this digital medium. Politicians and journalists on Twitter have the potential to shape public discourse and influence opinion in real-time, making it imperative to investigate their behavior in this digital space. Furthermore, addressing questions of gender bias in interaction frequency and content is crucial, as gender disparities in online political communication can have lasting implications for gender equality in Indian politics and journalism.

Although research has illuminated the behavior of politicians and journalists in individualistic societies (e.g., in global north), little attention has been given to understanding how collectivist societies, such as India, navigate these online spaces. Investigating the interaction frequencies and content between Indian politicians and journalists on Twitter from a gender perspective can help fill this gap and contribute to a more comprehensive understanding of the cultural nuances and gender dynamics at play in collectivist societies in the digital age.

This paper brings together several strands of the literature. Research shows that gender bias is widespread on social media, especially Twitter. Responses to tweets by female politicians are more likely to focus on their gender attributes than their professional and political acumen~\cite{paper-number-25-mertens-websci}. Researchers have also found the prevalence of gender stereotypes in two-way conversations between voters and politicians on Twitter~\cite{paper-number-8-evans, paper-number-9-evans, paper-number-13-meeks, paper-number-17-wagner}. These issues are compounded by the fact that women are structurally underrepresented in social media~\cite{paper-number-17-wagner, paper-number-26-gloor2015cultural}. In general female users suffer in their reach on Twitter because they have fewer followers and are retweeted less compared to men, which is investigated by Matias et al. by a nudge-based system called FollowBias~\cite{paper-number-12-nathan}. However, this work does nor look into popular users, i.e., female politicians who are in a significantly more prominent position than a general female Twitter user---we fill this gap. In earlier work on global north, the behavior of journalists on Twitter also seems to be divided on the lines of gender. Female political journalists interact more with other women on Twitter~\cite{paper-number-16-usher} and their behavior is different from that of male journalists~\cite{paper-number-14-parmelee, paper-number-15-parmelee}. In contrary, we show that in Indian Twitter both male and female journalists interacted more with male politicians compared to female politicians.  

\vspace{1mm}

\noindent \textbf{Summary:} Previous studies have shed light on gender bias in Twitter against female politicians and the divergence in the behavior of male and female journalists on Twitter. Howevere there is a gap in our understanding of gender bias in journalist-politician interaction in Indian Twitter. Specifically, we do not know if journalists promote or combat gender bias against politicians on Twitter. This paper attempts to fill this gap in the literature by specifically studying the tweets of journalists directed at the most popular male and female Indian politicians for gender bias.

\section{Data Collection}\label{sec:data}

\noindent In this section we describe our data collection process from Twitter. We specifically collected data about the interactions between specific Indian politicians and journalists on Twitter sampled  based on their popularity and gender. First, we start with how we created a list of Indian journalists and politicians for our study.

\subsection{Identifying Twitter accounts of Indian politicians and journalists}

\noindent \textbf{Identifying Twitter accounts of individual Indian politicians:} We leveraged a dataset of Indian Politicians from previous research by Pal et al.\cite{pal-2020-politicandatabase}. This dataset contained names and handles of multiple Indian Twitter accounts which are involved in politics (labelled as politicians). However, we noted that this dataset contained accounts of both political organizations (e.g., BJP for Andaman and Nicobar Islands) as well as individuals. To that end, we first cleaned the dataset, by cross-matching the names from this dataset with names from \texttt{MyNeta}\footnote{\url{https://www.myneta.info/}} which is an open data repository platform run by Association for Democratic Reforms (ADR) for bringing transparency to Indian elections. For each of the Indian political accounts in Pal et al. 's dataset, we searched \texttt{MyNeta} platform  with the name of the account. If the search found no politicians with this name, then we discard the account from our analysis as that account is probably not from an individual. At the end of the procedure, we ended up with 4,484  Twitter accounts of politicians.

\vspace{1mm}

\noindent \textbf{Identifying Twitter accounts of individual Indian political journalists:} Next, we focus on the Twitter accounts marked as individual journalists from a dataset of Twitter influencers released by Pal et al.'s previous research~\cite{pal-2020-journalistdatabase} (separate from accounts of media houses). There were 4,099 such accounts. However, we again faced a challenge---how can we identify the \textit{political journalists}? Specifically, we noted that this list contains several journalists who are not associated with political reporting and focus on areas such as entertainment, sports etc. Thus, we set to identify political journalists---journalist accounts that directly mentioned politicians' accounts in a non-trivial tweet (e.g., after discounting tweets with only emojis, urls, birthday greetings). To that end, we collected all the tweets posted by these 4,099 accounts between Jan 2020 and Dec 2022 using an open-source tool called crape. Then we discounted  tweets with only emoji, urls, greetings and checked if any of the final tweets mentioned an individual Indian politician's Twitter account (collected as described above). Finally, we include 3,214 journalists' accounts (78.4\%) in our dataset as political journalists. 

\vspace{1mm}

\noindent \textbf{Verifying the accuracy of Twitter accounts:} Finally, we manually verified if our filtering approach actually identified the correct Twitter accounts of Indian politicians and political journalists. We randomly sampled forty politicians and twenty journalist accounts. Then an author visited the actual Twitter accounts and read the first 20 tweets to ensure the account indeed belonged to an Indian politician (or political journalist). In 92.5\% of the random sample, our filtering approach correctly identified Twitter accounts of Indian politicians (or political journalists).

\subsection{Inferring gender of Indian politicians and political journalists}

\noindent Next, we infer the gender of the Twitter accounts of Indian politicians (or political journalists) as identified in the previous section. For this purpose, we used a service called \textit{Generize}~\cite{gender}. This service maps names to genders, is customized to Indian names, and previous studies reported high accuracy of gender inference from this service~\cite{generize-accuracy}. Once we infer gender of all accounts, for this study we focused on the most popular (by number of followers) politician and journalist accounts. Specifically, we sorted the politician accounts by the follower count and identified the top 50 accounts for male politicians and female politicians (as identified by Genderize). We further manually verified the accuracy of the inferred gender for these 100 Twitter accounts. We similarly identified the most popular 100 journalist accounts (50 male and 50 female).

\begin{table}[]
\centering
\footnotesize
\begin{tabular}{p{3cm}|lllll|}
\cline{2-6}
                                                                                                                             & \multicolumn{5}{c|}{\#Tweets posted by journalists}                                                                       \\ \cline{2-6} 
                                                                                                                             & \multicolumn{1}{l|}{Tot.}   & \multicolumn{1}{l|}{min} & \multicolumn{1}{l|}{avg.}  & \multicolumn{1}{l|}{median} & max   \\ \hline
\multicolumn{1}{|l|}{\begin{tabular}[c]{@{}l@{}}Male Journalists $\rightarrow$ Male politicians \\ \textbf{(MJ - MP)}\end{tabular}}   & \multicolumn{1}{l|}{10,032} & \multicolumn{1}{l|}{2}   & \multicolumn{1}{l|}{209}   & \multicolumn{1}{l|}{55}     & 1,563 \\ \hline
\multicolumn{1}{|l|}{\begin{tabular}[c]{@{}l@{}}Male Journalists $\rightarrow$ Female politicians \\ \textbf{(MJ - FP)}\end{tabular}} & \multicolumn{1}{l|}{1,293}  & \multicolumn{1}{l|}{1}   & \multicolumn{1}{l|}{34.0}  & \multicolumn{1}{l|}{13}     & 185   \\ \hline
\multicolumn{1}{|l|}{\begin{tabular}[c]{@{}l@{}}Female Journalists $\rightarrow$ Male politicians \\ \textbf{(FJ - MP)}\end{tabular}} & \multicolumn{1}{l|}{8,780}  & \multicolumn{1}{l|}{1}   & \multicolumn{1}{l|}{179.2} & \multicolumn{1}{l|}{49}     & 3,711 \\ \hline
\multicolumn{1}{|l|}{\begin{tabular}[c]{@{}l@{}}Female Journalists $\rightarrow$ Male politicians \\ \textbf{(FJ - FP)}\end{tabular}} & \multicolumn{1}{l|}{1,545}  & \multicolumn{1}{l|}{1}   & \multicolumn{1}{l|}{34.3}  & \multicolumn{1}{l|}{11}     & 482   \\ \hline
\end{tabular}
\caption{The \# of tweets posted by Indian journalists mentioning politicans. The Female politicians received relatively less mentioned tweets. }\label{tab:numbertweets}
\end{table}

\begin{table*}[]
\centering
\footnotesize
\begin{tabular}{l|p{12cm}|}
\cline{2-2}
                                                                                                                                              & \textbf{Sample tweet excerpts}                                                                                                                                                                                                 \\ \hline
\multicolumn{1}{|l|}{\multirow{2}{*}{\begin{tabular}[c]{@{}l@{}}Male Journalists $\rightarrow$ Male politicians \\ (MJ - MP)\end{tabular}}}   & Home Minister @[user] speaks to me in his first interview this election season in which he shares his UP poll prediction, speaks on the hijab controversy, Kashmir polls and a lot more.                   \\ \cline{2-2} 
\multicolumn{1}{|l|}{}                                                                                                                        & It is almost impossible to get any changes made in the Aadhar card despite doing all the things right ! What is a common man to do ? @[user]                                                               \\ \hline\hline
\multicolumn{1}{|l|}{\multirow{2}{*}{\begin{tabular}[c]{@{}l@{}}Male Journalists $\rightarrow$ Female politicians \\ (MJ - FP)\end{tabular}}} & Ideally \@[user] should have set this example ...but I hope all political leaders follow this example and cancel their political rallies                                                                    \\ \cline{2-2} 
\multicolumn{1}{|l|}{}                                                                                                                        & I wish to exhort all state governments, especially the states where reduction wasn't done during the last round (November 2021), to also implement a similar cut and give relief to the common man @[user] \\ \hline\hline
\multicolumn{1}{|l|}{\multirow{2}{*}{\begin{tabular}[c]{@{}l@{}}Female Journalists $\rightarrow$ Male politicians \\ (FJ - MP)\end{tabular}}} & Union transport minister @[user] too says public transport will resume soon with protocols and guidelines                                                                                                  \\ \cline{2-2} 
\multicolumn{1}{|l|}{}                                                                                                                        & I am all for fines for speeding in the very short burst where it is possible. But shouldn't Govt collecting tolls also ensure no loooong jams to pay tolls? @[user]                                        \\ \hline\hline
\multicolumn{1}{|l|}{\multirow{2}{*}{\begin{tabular}[c]{@{}l@{}}Female Journalists $\rightarrow$ Male politicians \\ (FJ - FP)\end{tabular}}} & Such welcome graciousness -- it would have made us all feel even better if we weren't seeing empty benches, with OUR ELECTED reps NEVER ON THE JOB or around even to APPLAUD!! @[user]                     \\ \cline{2-2} 
\multicolumn{1}{|l|}{}                                                                                                                        & does this make anyone angry or not? Owed Rs5000 crore. Offered Rs 500 crore. Paid Rs 5 crore upfront. Now take a bicycle loan and see how banks treat you @[user]                                          \\ \hline
\end{tabular}
\caption{Sample excerpts of \# of tweets posted by journalists mentioning politicians. We show tweets from four different categories based on the genders of sender and receiver. }\label{tab:sampletweets}
\end{table*}

\subsection{Collecting journalist-politician Twitter interaction data} 

\noindent Finally, to answer our research questions, we collect interaction data between the Indian politicians and political journalists' accounts. Specifically, we collected \textit{all} the tweets posted by 100 popular political journalist accounts and then filtered out the tweets which mentioned any of the 100 popular Indian politicians in our dataset. Thus, we divided our collected tweets into the following four categories---Male journalist’s tweets mentioning Male Politicians (\textbf{MJ-MP}), Female journalist’s tweets mentioning Male Politicians (\textbf{FJ-MP}), Male journalist’s tweets mentioning Female Politicians (\textbf{MJ-FP}) and Female journalist’s mentioning Female Politicians (\textbf{FJ-FP}). In total we collected 21,188 unique tweets. Note that a single tweet can mention multiple accounts. 

We note that, almost all hundred journalists across genders collectively mentioned our chosen popular politician accounts in their tweets. Furthermore, Table~\ref{tab:numbertweets} presents the number of tweets across our four categories. Notably, the accounts of female politicians received considerably less mention from both male and female Indian journalists. Now, we analyzed this interaction data collected from Twitter to identify potential gender bias in the journalist-politician interactions in Indian Twitter. Furthermore, Table~\ref{tab:sampletweets} presents tweet excerpts from each of the four categories. These example demonstrate that many of the tweets in our dataset across different categories are related to policy decisions and general governance.

\section{Analysis methodology}\label{sec:meth}

\noindent In this work, we primarily performed quantitative analysis to uncover general bias in Indian journalist-politician interaction. Specifically, we performed frequency analysis using statistical testing, emotion analysis, and topic analysis of the collected tweets. 

\vspace{1mm}

\noindent \textbf{Statistical analysis of interaction frequency and popularity}: To analyze gender bias in the frequency of interaction between Journalists and politicians, we examine the popularity of tweets between journalists and politicians across genders of politicians. If journalists do indeed promote gender bias, we expected to see more tweets by male journalists mentioning male politicians compared to mentioning female politicians. We expect female journalists to tweet less or as much about male politicians as they do about female politicians. To check for gender bias in the traction that tweets from politicians receive from journalists, we also examine the popularity of tweets (via retweets, replies and likes for these tweets). We use the Kruskal-Wallis H-Test to determine if significant differences exist in different categories of tweets. Furthermore, we employ pairwise Mann-Whitney U-test to analyze the differences in popularity across our four categories in more granular detail.

\vspace{1mm}

\noindent \textbf{Emotion analysis}: To examine gender bias in the content to the tweets addressed to politicians by journalists, we employ emotion analysis of tweets (using TweetNLP, a cutting-edge large language model-based Twitter-specific multilingual emotion detection tool~\cite{camacho-collados-etal-2022-tweetnlp}) from  ``MJ-MP'', ``MJ-FP'', ``FJ-MP'' and ``FJ-FP'' categories. Specifically, we seek to test for differences in these tweets for expressions of ``Anger'', ``Joy'', ``Optimism'' and ``Sadness''. We use a Kruskal-Wallis H test to determine if significant differences exist in the emotion scores of tweets (along the four dimensions) from our categories. 

\vspace{1mm}

\noindent \textbf{Topic analysis}: Finally, we analyze the content of the tweets across the four categories. Specifically, we used a Latent Dirichlet Allocation (LDA) Topic detection model (from Gensim library) for performing LDA. We used a coherence score analysis approach where for each category we varied the number of topics from 2 to 15 and in each output, checked the coherence score (signifying the interpretability of the detected topics). When increasing the number of topics did not increase the coherence score, we found the optimal number of topics. We set the $\alpha$ and $\eta$ parameters as $\frac{1}{num\_topics}$ as identified in Gensim tutorials~\cite{gensim-tutorial}. The optimal number of topics for each category was thirteen.

\begin{figure*}[t]
\centering
\begin{subfigure}{0.24\textwidth}
    \includegraphics[width=\textwidth]{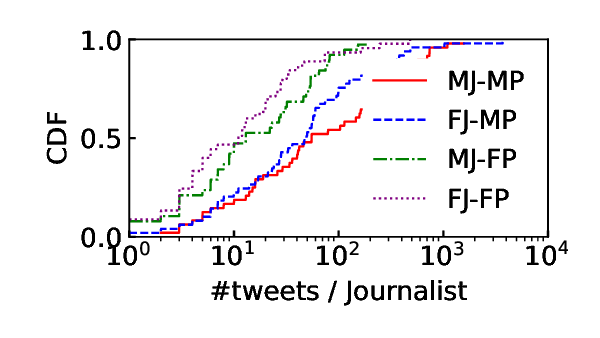}
    \caption{}
    \label{fig:first}
\end{subfigure}
\begin{subfigure}{0.24\textwidth}
    \includegraphics[width=\textwidth]{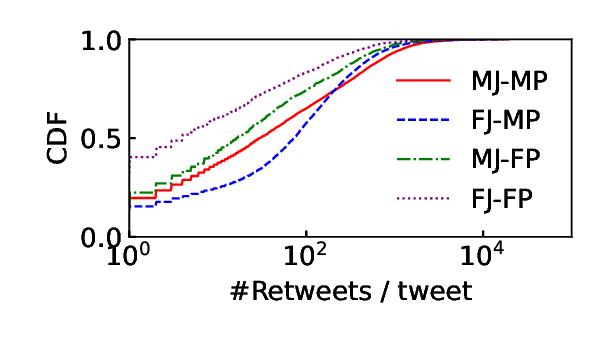}
    \caption{}
    \label{fig:second}
\end{subfigure}
\begin{subfigure}{0.24\textwidth}
    \includegraphics[width=\textwidth]{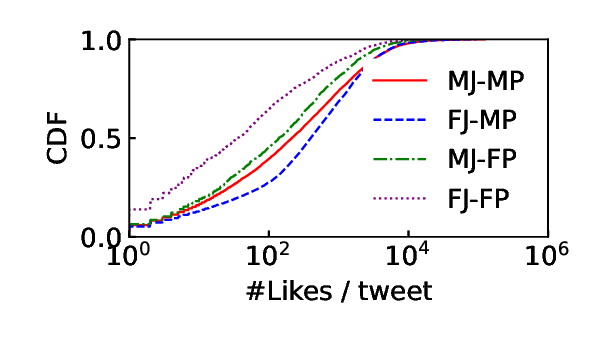}
    \caption{}
    \label{fig:third}
\end{subfigure}
\begin{subfigure}{0.24\textwidth}
    \includegraphics[width=\textwidth]{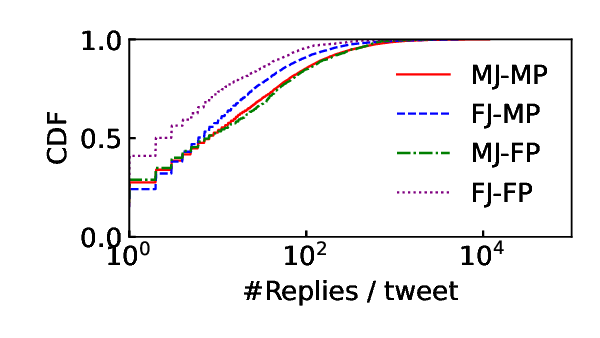}
    \caption{}
    \label{fig:fourth}
\end{subfigure}

\caption{The difference in interaction frequency and popularity of tweets from four categories: Male journalist $\rightarrow$ Male politician (MJ-MP), Female journalist $\rightarrow$ Male politician (FJ-MP), Male journalist $\rightarrow$ female politician (MJ-FP) and Female journalist $\rightarrow$ Female politician (FJ-FP). Figure~\ref{fig:first} presents the \# tweets per journalist in each category. Figure~\ref{fig:second}, Figure~\ref{fig:third} and Figure~\ref{fig:fourth} presents a comparison of the popularity of tweets as captured by \#retweets, \#likes and \#replies received by the tweets. Female politicians received less interaction (i.e., mentioned tweets) from journalists and these interactions are less popular compared to male politicians.} 
\label{fig:interaction}
\end{figure*}

\section{Results}\label{sec:results}

\subsection{Gender bias in interaction frequency and popularity of journalist-politician interactions (RQ1)}

\noindent In order to explore the answer to the first research question, we started with checking if there is a gender bias in the frequency of interaction (i.e., frequency of mentions) between politicians and journalists on Twitter. 

\vspace{1mm}

\noindent \textbf{Male Politicians are more frequently mentioned by journalists}: Figure~\ref{fig:first} compares the CDF of the number tweets posted by journalists mentioning male and female politicians. We make an interesting observation from this figure. When the receiving politician is male (i.e., in MJ-MP and FJ-MP categories), the number of mentioned tweets (and hence the frequency of journalist-politician interaction) is higher when female politicians are at the receiving end. To that end, a Kruskal-Wallis test among the number of tweets per journalist within the four categories revealed quite statistically significant differences across categories ($p << 0.05 $). Then we performed pairwise Mann-Whitney tests for following up among the four categories (MJ-MP, MJ-FP, FJ-MP, FJ-FP). There is no statistically significant difference when either a Male or Female journalist mentions Male politician accounts. Similarly, there is no statistically significant difference when either a Male or Female journalist mentions a Female politician account. However, there are statistically significant differences in how frequently Male/Female journalists mention Male politicians vs. how frequently they mention Female politicians (all $p << 0.05$). Next we compare the popularity per tweet directed toward Male vs. Female politicians. 
  
\vspace{1mm}

\noindent \textbf{Tweets mentioning Male politicians are more popular:} Figure~\ref{fig:second}, Figure~\ref{fig:third} and Figure~\ref{fig:fourth} compares the popularity of tweets from different categories as captured by number of retweets, likes and replies received by individual tweets. These plots indicate that across all types of popularity measures, the tweets posted by Male journalists mentioning Male politicians (MJ-MP) are more popular than the tweets posted by Female journalists mentioning Female politicians (FJ-FP). To that end, a kruskal-wallis test indicated that indeed  popularity of the tweets are statistically significantly different across four categories ($p$ ranged from $8.7 \times 10^{-56}$ to $1.3\times 10^{-185}$). In a follow-up analysis, for each of these popularity measures and for each of the four categories we performed pairwise Mann-Whitney U-test across the four categories (MJ-MP, MJ-FP, FJ-MP, FJ-FP). We found that, in all the cases, the differences are statistically significant. In fact, across all metrics (retweets, likes, replies) the median popularity of the posts mentioning female politicians is consistently lower than posts mentioning male politicians. E.g., the median number of likes received by a post in FJ-FP is 35 and the median number of likes received by a post in FJ-MP is 398---more than ten times higher. Moreover, tweets on female politicians by male journalists receive almost four times more ``Likes'' than tweets on female politicians by female journalists (median 139 likes vs, median 35 likes respectively). 

Our observation implies that Twitter users in India seem to ascribe greater credibility to the views on male journalists on female politicians compared to the views of female journalists on female politicians. These observations hold for ``retweets'' as well. Overall, our popularity analysis of these four categories of tweets reveals that while journalists do not suffer from explicit bias in their interactions with politicians, there is evidence to support the existence of gender bias in the amount of interest these interactions generate from active Twitter users.    

\subsection{Gender bias within the content of journalist-politician tweets  (RQ2)}

\noindent In the last section, our analysis showed a significant bias towards the Male politicians from both male and female journalists---the tweets mentioning male politicians are more frequent as well as more popular. However, to that end, we checked if the content of these tweets might be responsible for this bias. Specifically, we checked the emotion and the topic of tweets written by male / female journalists and directed towards male / female politicians. 

\subsubsection{Emotion analysis}: We used the TweetNLP tool for detecting the emotions of the tweets for each category~\cite{camacho-collados-etal-2022-tweetnlp}. TweetNLP provides a diachronic large-language model (TimeLMs) based approach for detecting emotion, specifically from multi-lingual tweets.  The goal of this analysis is to determine if there are significant differences in the emotional scores of tweets---if there is that might indicate a gender bias inherently in the tweets based on the gender of sender and receiver. We considered four main emotions: anger, joy, optimism, and sadness and each tweet in each of the four categories were assigned for emotion score along these dimensions. Then we performed a Kruskal-Wallis test to identify if any of the emotions were different across the four categories (MJ-MP, MJ-FP, FJ-MP, FJ-FP). We found that the $p$-value for each of the four tests (one for each emotion) ranged from $0.16$ to $0.99$, hinting at no statistically significant difference within the emotions of the tweets.

\subsubsection{Topic analysis}: To dig further, we performed a topic analysis of the tweets (using Latent Dirichlet Allocation or LDA) collected across four categories. The goal was to check if the topics of the tweets changed based on the gender of the sender or receiver. As described in Section~\ref{sec:meth} we identified the optimal number of topics (which are essentially clusters of words) for each category and identified the most significant five words for each topic using the LDA algorithm. For each of the four categories, the optimal number of topics were thirteen. Next, we identified the thirteen topics using the LDA algorithm for each category of tweets and performed a significant word analysis for the detected topics. Specifically, for each category of tweets we selected the topics (e.g., topics from MJ-MP) and picked the significant words representing each topic. Then for each topic we checked if those words also occured in the topics detected from other categories of tweets (if found it will signify that words representing topics are also present in topics detected from other categories of tweets). For each of the four categories of tweets on average 81.5\% to 93.8\% significant words (representing the topics) occur in topics detected from tweets of other categories. 

This analysis supports our observation from emotion analysis---the content of tweets across those four categories are the same. However, still the tweets directed towards male politicians attract more interaction compared to the tweets directed towards female politicians. Next we explore a potential reason for this gender bias.

\subsection{Potential reason for the gender bias}

\subsubsection{Inherent Gender Bias in Indian Twitter:} We checked a simple statistic regarding top politicians---how many of the most popular politicians (based on the number of Twitter followers) are male vs. they are female. To that end, we leveraged our dataset of top politicians and checked the gender of top 85 politicians (whose Twitter accounts are also part of this study). This analysis uncovered an unsettling gender imbalance among the top politicians---out of 85 top politicians 58 are male and 26 are female. Thus, popular male politicians are almost twice in number compared to popular female politicians. We postulate that this inequality is one of the key reasons behind our observed phenomenon of male politicians attracting significantly more interaction from the general public as well as journalists. 

In fact, this inequality reflects a systemic bias deeply ingrained in society. This gender disparity extends its influence even to the realm of Twitter, where male politicians tend to garner a larger number of followers than their female counterparts. This phenomenon isn't isolated; it permeates various sectors, as illustrated by the dominance of men in top positions across industries. In corporate boardrooms, technology firms, and the entertainment sector, leadership roles are predominantly occupied by men. This systematic bias, rooted in societal norms, is further reinforced by the strong correlation between social capital and the attainment of positions of power. Consequently, popularity on Twitter serve as a stark reflection of this intrinsic bias. Addressing these disparities is paramount for fostering gender equality and dismantling deeply entrenched biases in society.

\section{Limitations}\label{sec:lim}

\noindent This study focuses only on the top female/male journalists (based on their follower count) on Twitter. These journalists work for reputable media outlets which likely have strong guidelines and policies on gender. Therefore, our results underestimate the gender bias against politicians on Twitter. A more granular analysis of the tweets of lesser known journalists and their followers may reveal even higher levels of bias and misogyny. A similar assessment can be made on our choice of politician. We focus on the top 50 politicians of each gender in our analysis. These politicians are of considerable standing and thus less likely to face bias in their interactions with journalists. This may not be true in situations where journalists interact with female politicians of lower standing and political clout.  Thus by focusing on the top fifty female politicians, we underestimate the bias that average female politicians may be subjected to on Twitter. Lastly, our analysis also reveals that there might be bias against female journalists on Twitter. This study does not comprehensively address this, especially in the context of their interactions with male and female politicians. We expect to address this in future work.

\section{Concluding Discussion}\label{sec:conclu}

\noindent Misogyny remains an intractable social problem in India and the subcontinent. Subconscious gender bias manifests itself in many spheres of life and in several professions. This study seeks to uncover gender bias against female politicians on Twitter in India using NLP-based emotion analysis and LDA topic modeling- specifically in the interactions between journalists, who are the main drivers of political discourse on Twitter, and politicians.  

Our findings reveal a disconcerting trend: while popular journalists themselves do not seem to suffer from gender bias, on average, Twitter users display bias against female politicians. This bias is a reflection of the larger societal issue where male politicians traditionally occupy a higher pedestal in politics. This systematic bias is translating onto Twitter, and merely using balanced language is not enough to rectify the situation.

To counteract this bias, we suggest utilizing recommender systems to amplify the voices of female politicians, as balanced language alone isn't sufficient.
Additionally, our analysis reveals that users tend to give more weight to the views of male journalists when evaluating female politicians, highlighting the need for diverse voices in political discourse. 

In summary, this study emphasizes the need to address gender bias on Twitter through not only language but also platform design and algorithms. Creating a more inclusive digital space is crucial to ensuring female public figures can effectively disseminate their views despite societal biases.

\bibliographystyle{ACM-Reference-Format}
\bibliography{genderbias}

\end{document}